\documentclass[aps,twocolumn,prl]{revtex4}
\usepackage{times}
\usepackage{graphicx}% Include figure files
\usepackage{dcolumn}% Align table columns on decimal point
\usepackage{bm}% bold math
\usepackage{amsmath}
\usepackage{color,psfrag,epsfig}
\begin{document}

\newcommand{\bk}{{\bf k}}
\newcommand{\bc}{\begin{center}}
\newcommand{\ec}{\end{center}}
\newcommand{\mub}{{\mu_{\rm B}}}
\newcommand{\sD}{{\scriptscriptstyle D}}
\newcommand{\sF}{{\scriptscriptstyle F}}
\newcommand{\sCF}{{\scriptscriptstyle \mathrm{CF}}}
\newcommand{\sH}{{\scriptscriptstyle H}}
\newcommand{\sAL}{{\scriptscriptstyle \mathrm{AL}}}
\newcommand{\sMT}{{\scriptscriptstyle \mathrm{MT}}}
\newcommand{\sT}{{\scriptscriptstyle T}}
\newcommand{\up}{{\mid \uparrow \rangle}}
\newcommand{\down}{{\mid \downarrow \rangle}}
\newcommand{\upsp}{{\mid \uparrow_s \rangle}}
\newcommand{\downsp}{{\mid \downarrow_s \rangle}}
\newcommand{\upsone}{{\mid \uparrow_{s-1} \rangle}}
\newcommand{\downsone}{{\mid \downarrow_{s-1} \rangle}}
\newcommand{\upt}{{ \langle \uparrow \mid}}
\newcommand{\downt}{{\langle \downarrow \mid}}
\newcommand{\bbar}{{\mid \uparrow, 7/2 \rangle}}
\newcommand{\abar}{{\mid \downarrow, 7/2 \rangle}}
\renewcommand{\a}{{\mid \uparrow, -7/2 \rangle}}
\renewcommand{\b}{{\mid \downarrow, -7/2 \rangle}}
\newcommand{\plus}{{\mid + \rangle}}
\newcommand{\minus}{{\mid - \rangle}}
\newcommand{\psio}{{\mid \psi_o \rangle}}
\newcommand{\psis}{{\mid \psi \rangle}}
\newcommand{\bpsio}{{\langle \psi_o \mid}}
\newcommand{\barpsi}{{\mid \psi' \rangle}}
\newcommand{\barpsio}{{\mid \bar{\psi_o} \rangle}}
\newcommand{\ex}{{\mid \Gamma_2^l \rangle}}
\newcommand{\LH}{{{\rm LiHoF_4}}}
\newcommand{\LHx}{{{\rm LiHo_xY_{1-x}F_4}}}
\newcommand{\de}{{{\delta E}}}
\newcommand{\Ht}{{{H_t}}}

\title{Quantum spin glass and the dipolar interaction}
%\title{Absence of spin-glass long range order for random

\author{Moshe Schechter}
\author{Nicolas Laflorencie}
\affiliation{Department of Physics \& Astronomy, University of British Columbia,
Vancouver, B.C., Canada, V6T 1Z1}
\date{}

\begin{abstract}

Systems in which the dipolar energy dominates the magnetic
interaction, and the crystal field generates strong anisotropy
favoring the longitudinal interaction terms, are considered. Such
systems in external magnetic field are expected to be a good
experimental realization of the transverse field Ising model. With
random interactions this model yields a spin glass to paramagnet
phase transition as function of the transverse field. Here we show
that the off-diagonal dipolar interaction, although effectively
reduced, destroys the spin glass order at any finite transverse
field. Moreover, the resulting correlation length is shown to be
small near the crossover to the paramagnetic phase, in agreement
with the behavior of the nonlinear susceptibility in the experiments
on $\LHx$. Thus, we argue that the in these experiments a cross-over
to the paramagnetic phase, and not quantum criticality, was
observed.

\end{abstract}

\maketitle

The study of quantum phase transitions (QPT) is of prime recent
interest, as it is believed that the understanding of the physics at
the vicinity of quantum critical points will shed light on some of
the most interesting problems in condensed matter physics, such as
the metal insulator transition, superconducting insulator transition
and high temperature superconductivity. Quantum magnets, and
specifically their modeling by the transverse field Ising model
(TFIM)

\begin{equation}
  H =  - \sum_{i,j} J_{ij} \tau_i^z  \tau_j^z -
\Delta \sum_i \tau_i^x \, .
 \label{IsingH}
\end{equation}
are a particularly good laboratory to study QPT, as this model is
rich enough to capture the interesting physics of QPT, yet simple
enough to allow theoretical treatment. Experimentally, much effort
was invested to realize the TFIM, and maybe the best realization is
in anisotropic dipolar systems. In these systems the dipolar energy
dominates the spin-spin interaction, and the crystal field generates
strong anisotropy resulting in a ground state Ising like doublet for
the single spins and an effective reduction of all but the
longitudinal interaction terms.

Indeed, $\LHx$ with x$=1$ was shown~\cite{BRA96} to exhibit a
ferromagnetic to paramagnetic (PM) transition as function of
transverse field $\Ht$ and temperature $T$.  As x is reduced, the
randomness in the position of the magnetic Ho atoms results in
frustration, and for x$=0.167$ a spin-glass (SG) phase was
observed~\cite{WER+91,WBRA93}. Furthermore, applying a transverse
magnetic field induces quantum fluctuations, leading to a PM phase
at large fields. Thus, this compound is considered to be the
archetypal experimental realization of a quantum
SG~\cite{CDS96,KR04}.

\begin{figure}[ht!]
\psfrag{X}{$\propto H_{t}^{-\nu}$}
\includegraphics[width=5cm,clip]{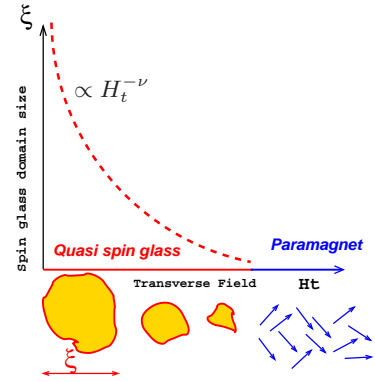}
\caption{Schematic picture for the $T=0$ behavior of a dipolar Ising
SG in a transverse field $H_t$. The typical size $\xi$ of a SG
ordered domain (depicted below the x-axis) decreases with $H_t$,
with a critical exponent $\nu$ (see text). At large enough $H_t$ the
system becomes PM, {\it via a crossover and not a quantum critical
point}.} \label{fig:PhDg}
%
%This low field phase, called {\it{Quasi spin glass}}, is destroyed
%for large enough $H_t$ when $\xi$ is of order unity and becomes a
%paramagnet.} \label{fig:PhDg}
\end{figure}

In this letter we show that for anisotropic dipolar glasses in
general, and for the $\LHx$ compound in particular, the off-diagonal
terms of the dipolar interaction, albeit effectively reduced,
qualitatively change the physics of the problem.  In particular, in
the presence of a transverse field the off-diagonal dipolar terms
reduce the symmetry of the system in comparison to the TFIM, and
render the latter inadequate in studying the system. A proper
treatment of the off-diagonal dipolar terms results in the absence
of long-range SG order at any finite $\Ht$, and a reduction of the
cusp in the non-linear susceptibility at the crossover to the PM
phase. Thus, we argue that the experimental line drawn at the peak
values of the non-linear susceptibility\cite{WBRA93} is not a phase
transition line. Except for the point at $\Ht = 0$, this line
corresponds to a cross-over between a paramagnet to a phase we
denote a "quasi spin-glass" (QSG). In this phase the system
separates into domains within which the random ordering of the spins
is maintained. These domains have a typical size $\xi(H_t)$ which
dictates the correlation length in the system, and its dependence on
$H_t$ is given by the critical exponent $\nu$ calculated below. The
domain structure is maintained until the crossover field, where
fluctuations between the relevant Ising like states dominate and the
system becomes PM (see Fig.\ref{fig:PhDg} for a schematic picture at
$T=0$). This crossover is expressed as a cusp in the non-linear
susceptibility. Importantly, the reduction of $\xi$ with increasing
$\Ht$ results in the corresponding reduction of the cusp in the
nonlinear susceptibility, explaining the peculiar experimental
result\cite{WBRA93} where the cusp is reduced {\it with decreasing
$T$}. Interestingly, we show below that at $T=0$ the crossover takes
place at a value of $\Ht$ which corresponds to $\xi \approx 1$, and
therefore to a complete absence of a cusp of the non-linear
susceptibility at $T=0$, as can be inferred from the
experiment\cite{WBRA93}.

{\it Theoretical considerations.---} Our analysis below is valid
both specifically to the $\LHx$ system, as we further comment on
below, as well as to any anisotropic dipolar system. The only
requirement is that
%Since our analysis below relies only on
the single spins have a ground state Ising like doublet, with a
large energy separation to the excited states. To emphasize the
generality of our approach we consider the following spin-$s$
Hamiltonian
\begin{equation}
{\cal{H}} = - D \sum_i [(S_i^z)^2 - s^2] -
\frac{1}{2} \sum_{i \neq j,\alpha, \beta} V_{ij}^{\alpha \beta}
S_i^\alpha  S_j^\beta - \mub H_t \sum_i S_i^x \, .
 \label{genH}
\end{equation}
Here $i,j$ denote the positions of the spins, randomly diluted on some
lattice, $V_{ij}^{\alpha \beta}$ denotes the dipolar interaction
($\alpha, \beta = x,y,z$), and $D>0$ is the anisotropy constant due to
the crystal field.  For $H_t=0$ the GS of a single spin is doubly
degenerate with $s_z=\pm s$ and zero energy.  The corresponding states
are denoted $\upsp$ and $\downsp$. The first excited states have
$s_z=\pm (s-1)$ and energy $\Omega_o=(2s-1)D$.  Throughout the paper
we assume that $\Omega_o \gg \mub H_t, V_{max}$ where $V_{max} $ is
the largest dipolar energy between two spins in the system.  We now
define ${\cal{H}} = {\cal{H}}_{\|} + {\cal{H}}_{\perp}$ such that
\begin{equation}
{\cal{H}}_{\|} = - D \sum_i [(S_i^z)^2 - s^2] - \frac{1}{2}
\sum_{i \neq j} V_{ij}^{zz} S_i^z  S_j^z,
\label{DHo}
\end{equation}
and
\begin{equation}
{\cal{H}}_{\perp} = - \frac{1}{2}
\sum_{i \neq j} \sum_{(\alpha \beta)\neq(zz)}
V_{ij}^{\alpha \beta}
S_i^\alpha  S_j^\beta - \mub H_t \sum_i S_i^x \, .
\label{DHint}
\end{equation}
We assume that the dilution is such that ${\cal{H}}_{\|}$ is
equivalent to the classical Ising model with random interactions and
exhibits a SG phase at low temperature. As this classical dipolar
Ising SG is equivalent to the short range Edwards Anderson
model~\cite{BMY86,SL06} (Eq.(\ref{IsingH}) with random nearest
neighbor $J_{ij}$ and $\Delta=0$~\cite{EA75}), our analysis is done
within the scaling (``droplet'') picture~\cite{FH86} which accounts
for its behavior at large sizes. The GS of ${\cal{H}}_{\|}$ is then
two fold degenerate~\cite{FH86} with states $\psio, \barpsio$, which
are related by $S_z \rightarrow -S_z$ symmetry, and in which each
spin is in either state $\upsp$ or $\downsp$. Importantly, adding a
transverse field term preserves the above symmetry, and therefore
the TFIM is the archetypal model for the quantum SG phase. However,
when adding ${\cal{H}}_{\perp}$ {\it which includes the off-diagonal
dipolar terms, this symmetry is not preserved}. The GS degeneracy
breaks, and the system gains energy by choosing locally a state
similar to $\psio$ or $\barpsio$ according to which optimizes the
energy gain due to ${\cal{H}}_{\perp}$.

Following the scaling picture of Fisher and Huse~\cite{FH86} and
using an Imry-Ma~\cite{IM75} like argument we calculate this energy
gain, i.e. the energy to flip a droplet of size $L$ having $N \sim
L^d$ spins, due to the addition of ${\cal{H}}_{\perp}$. This energy
gain (see Eq.(\ref{deq}) below) is then compared with the energy
cost due to the domain wall formation, and the correlation length is
obtained (\ref{corl}). Although, as is shown below, one can define
an effective longitudinal random field at each site and use a direct
analogy to the Ising SG in a random field, we would proceed by
calculating directly the energy gain.

Consider first
\begin{equation}
  {\cal{H}}_{\perp}^{'} = - \sum_{i \neq j} V_{ij}^{zx}
S_i^z  S_j^x    - \mub H_t \sum_i S_i^x \, .
 \label{intHp}
\end{equation}
The addition of ${\cal{H}}_{\perp}^{'}$ to ${\cal{H}}_{\|}$ changes
$\psio \rightarrow \psis$ and $\barpsio \rightarrow \barpsi$ with
energies $E_{\psi}$ and $E_{\psi^{'}}$ respectively. The energy the
system gains by choosing locally the lowest energy state is $\delta E
\equiv | E_{\psi} - E_{\psi^{'}}|$ which we now calculate. In second
order perturbation theory $ E_{\psi} = E_{\psi_o} + E_{\psi}^{(2)}$
where
\begin{equation}
 E_{\psi}^{(2)} =  - \frac{\bpsio ( \sum_{i \neq j} V_{ij}^{zx}
S_i^z  S_j^x + \mub H_t \sum_i S_i^x)^2 \psio}{\Omega_o} \, .
 \label{E2}
\end{equation}
Here we used the fact that the only relevant excited states are those
in which one spin changes its state from $s_z = \pm s$ to $s_z = \pm
(s-1)$.  Therefore the energy of all relevant excited states is
$\Omega_o$ in leading order, and the sum over the excited states can
be taken out as the identity operator. A similar equation holds for
$E_{\psi^{'}}$. One can show that the terms with even powers of $H_t$
are equal for $E_{\psi}$ and $E_{\psi^{'}}$, while the term linear in
$H_t$ is equal in magnitude but has opposite signs for $E_{\psi}$ and
$E_{\psi^{'}}$~\cite{SL06}.  Using the fact that since $k \neq l$ the
operators commute we obtain
\begin{equation}
\delta E = \frac{4}{\Omega_o}
\bpsio \mub H_t \sum_i S_i^x \sum_{k \neq l}
V_{kl}^{zx} S_k^z  S_l^x \psio \,
 \label{E2n}
\end{equation}
and therefore
\begin{equation}
\delta E = 4 \frac{s \mub H_t}{2 \Omega_o} \sum_{k \neq i} V_{ki}^{zx}
\bpsio S_k^z  \psio =  \frac{2 s \mub H_t}{\Omega_o} \sum_i h_i^x  \, ,
 \label{de2f}
\end{equation}
where we define $h_i^x \equiv \sum_{k} V_{ki}^{zx} \langle S_k^z
\rangle$ as an effective transverse magnetic field at site $i$. For
each $i$ all the $V_{ki}$'s are small except the few for which the
sites $i$ and $k$ are spatially close. Due to the randomness of the
sign, retaining for each $i$ the term with the largest absolute
value, denoted $\tilde{V_i}$, gives a good approximation for $\delta
E$ up to a numerical factor $c$ of order unity. Since $\tilde{V_i}$
is random in sign, the average energy gained by flipping a droplet
of $N$ spins is given by
\begin{equation}
\langle \delta E \rangle= c \frac{s^2 \mub H_t V \sqrt{N}}{\Omega_o} \, ,
 \label{deq}
\end{equation}
where $V$ is the average magnitude of $\left| \tilde{V_i}\right|$, and
we choose $\psio$ and $\barpsio$ such that $\de > 0$.

The above result (\ref{deq}) is central to our analysis, and in
order to check our approximation of randomness leading to it we
calculated the gap between the GS and the first excited state
numerically using Lanczos exact diagonalization (ED)~\cite{ED}.  We
consider system sizes in the regime where they are much smaller than
$\xi$. This is important for our calculation, since then we are
dealing with single domains, and therefore the two lowest states
correspond to $\psis, \barpsi$, and the gap to $\de$. To reproduce
the experimental situation, we focus on three dimensional finite
size clusters, randomly distributing $N$ spins at the rare earth
sites of the $\LHx$ lattice.  Since we are interested in small
fields, it is sufficient to use $s=1$ particles with on-site
anisotropy $\Omega_o$. We therefore study the spin-1 version of
$\cal{H}=\cal{H}_{\|}+\cal{H}_{\perp}^{'}$:
\begin{eqnarray}
{\cal{H}}_{1} = - \sum_{i \neq j}\left[\frac{1}{2}V_{ij}^{zz}S_i^z
S_j^z
+V_{ij}^{zx}S_i^z  S_j^x\right]\nonumber \\
-\mub H_t\sum_iS_{i}^{x} - \Omega_o\sum_i\left([S_i^z]^2 -
s^2\right) \, . \label{s1}
\end{eqnarray}
Here and below all energy scales are expressed in units of the
typical n.n. dipolar energy $V_0$. We fixed the dilution to a
constant $x=3/16=18.75\%$ by using $2\times 2\times N/3$ unit cells,
$N$ being the total number of spins (there are $4$ rare earth sites
per unit cell).  For this dilution we find that $V=0.8$. Thus the
Lanczos ED have been performed in the {\it{full}} $d=[2s+1]^N$
dimensional Hilbert space for $N=6,~9$ and $12$ $s=1$
spins\cite{note2}. Then, the gaps have been computed for $10,000$
independent random samples for each size.

\begin{figure}[ht!]
%\bc
\psfrag{L}{$\ln \frac{\delta E}{\sqrt{N}}$}
\psfrag{P}{${\rm{P}}\left(\ln\frac{\delta E}{\sqrt{N}}\right)$}
\psfrag{B}{${\langle{\delta E\rangle}}$} \psfrag{A}{$\sqrt{N}$}
\epsfig{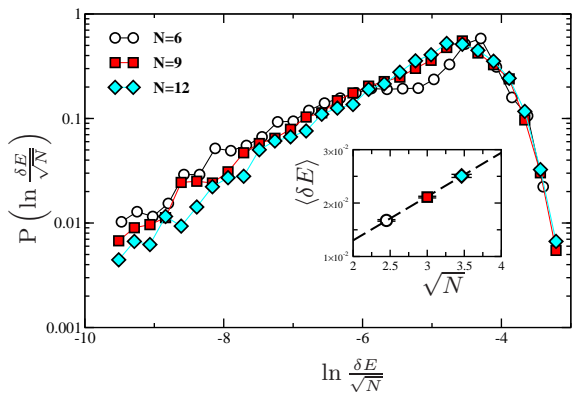}
\caption{Distribution P$\left(\ln\frac{\delta E}{\sqrt{N}}\right)$
    plotted in a semi-log scale.
      Lanczos ED data obtained for the spin-1 Hamiltonian
    (\ref{s1}) with $\Omega_o=50$ and $\mub H_t=0.5$, have been collected
    over 10,000 random
    samples with the $\LHx$ structure and a dilution $x=18.75\%$.
    Three different sizes have been used: $N=6,~9$ and
    $12$, as indicated on the plot. Inset: Linear
    dependence of the disorder average gap $\langle \delta E\rangle$
    vs $\sqrt{N}$. The dashed line is a fit of the form
    $\alpha\sqrt{N}$
    with $\alpha\simeq 0.008$.}
\label{fig:1}
%\ec
\end{figure}

In Fig.~\ref{fig:1} we present the numerical results obtained in the
perturbative regime (i.e. $\Omega_o\gg \mub H_t$ and $(\mub
H_t)^2/\Omega_o \ll V_{max}$). The $\sqrt{N}$ scaling of $\delta E$
as stated in Eq.~(\ref{deq}) is clearly demonstrated, as we found a
very good data collapse~\cite{note1} for the distribution of $\ln
\frac{\delta E}{\sqrt{N}}$.  The inset of Fig.~\ref{fig:1} also
shows that the disorder average gap $\langle \de \rangle
=\alpha\sqrt{N}$. Confronting the numerical estimate obtained for
the prefactor $\alpha$ with Eq.~(\ref{deq}), we get $c\simeq 1$.  We
have also checked the scaling of the gap with $H_t/\Omega_o$, where
for several combinations of ($\Omega_o$,$H_t$), we obtain an
excellent collapse of the data by rescaling $\delta E\to\frac{\delta
E}{H_t/\Omega_o}$~\cite{SL06}.

In order to obtain the correlation length of the system, i.e. the
typical domain size, we have to compare the domain's energy gain
\ref{deq} to the energy cost due to the formation of a domain wall.
For the short range Ising SG this energy is $\propto L^{\theta}$
with $\theta \approx 0.2$ in $3$ dimensions~\cite{BM84,McM84}.
Furthermore, under quite general conditions Fisher and Huse
argued~\cite{FH86} that {\protect{$\theta \leq (d-1)/2$.}}  For the
dipolar Ising SG we expect the same scaling behavior with a similar
exponent $\theta_d \simeq \theta$ to hold~\cite{BMY86,FH86,SL06},
and the energy of flipping a domain is therefore $\approx V s^2
L^{\theta_d}$.  As a result, for $L$ such that $(s^2 \mub H_t V
\sqrt{N})/\Omega_o > V s^2 L^{\theta_d}$ it will be preferable for
domains to choose their state between $\psis$ and $\barpsi$ as the
one that locally minimizes $E_{\psi}^{(2)}$. This results in a
finite correlation length
\begin{equation}
\xi \approx \left(\frac{\Omega_o}{\mub H_t}\right)^\frac{1}{(3/2) -
\theta_d} \, .
 \label{corl}
\end{equation}
For the Ising SG in longitudinal field it was
argued~\cite{FH86,FH87} and then verified
experimentally~\cite{MJN+95,JTKI05,SS05b} and
numerically~\cite{TH04,YK04} that there is no de Almeida Thouless
line~\cite{AT78}, and no SG phase at any finite field.  At $H_t \ll
\Omega_o/\mub$ our system is equivalent to the above model in small
longitudinal fields, and we thus argue that there is no SG phase at
any finite {\it transverse} field when the interaction is dipolar,
and as $H_t \rightarrow 0$ the correlation length diverges with the
same form~\cite{FH86} of the critical exponent $\nu = \frac{1}{(3/2)
- \theta_d}$.

In our treatment the only dipolar terms we considered are the
longitudinal and the ${zx}$ terms. However, one can show that all the
neglected terms [see the terms present in ${\cal{H}}_{\perp}$,
Eq.~(\ref{DHint}) but not in ${\cal{H}}_{\perp}^{'}$,
Eq.~(\ref{intHp})] do not contribute to $\delta E$ in second order
perturbation theory~\cite{SL06}.

Interestingly, the two effects of the transverse magnetic field,
i.e. inducing the crossover to the paramagnetic phase, and the
reduction in $\xi$ calculated above, behave very differently as
function of $\Ht$. The former is dictated by fluctuations between
the two single spin Ising ground states, which depend on $H_t$ to a
high power, of order $s$, and are practically negligible as long as
$\Ht \ll \Omega_0/\mub$. However, {\it the fluctuations that dictate
the reduction of $\xi$ at low transverse fields are between each
single spin ground state and its first excited state} at energy
$\Omega_0$. The latter depend on $H_t$ to second order and result in
a reduction of $\xi$ which depends on $1/H_t$ to a power $\nu$ close
to unity. Therefore, the disordering of the SG order by $H_t$ occurs
in two stages. At low field domains of size $\xi$ are formed, within
which the GS is very similar to either of the two zero field SG
ground states. At $H_t \approx \Omega_o/\mub$ a crossover occurs
where the order within each domain is destroyed, and each spin in
the system fluctuates independently. Importantly, when reaching the
crossover region at very low $T$ one is already in the regime where
$\xi \approx 1$ in units of inter-spin spacing, resulting in small
features in the relevant susceptibilities, in agreement with
experiment~\cite{WER+91,WBRA93}. We would like to emphasize that the
understanding of the scenario above {\it requires} a model in which
the large spins are considered and the anisotropy in explicitly
taken into account. Indeed, the anisotropy energy $\Omega_0$ enters
explicitly into Eqs.(\ref{deq}),(\ref{corl}). A presumably simpler
model in which one treats spin-half particles and models the
effective reduction of the off-diagonal terms in the dipolar
interaction by a multiplicative reduction factor will not be
sufficient, since in such a model both the reduction in the
correlation length and the crossover to the PM phase are induced by
fluctuations between the Ising ground states, and therefore have the
same scale in magnetic field.

In addition to changing the symmetry of the system at $H_t \neq 0$,
resulting in the destruction of the SG phase and the QPT to the PM
phase, the off-diagonal terms of the dipolar interaction also
enhance the effective transverse field\cite{SS05}. Although, in
principle, $h_i^x$ [see Eq.(\ref{de2f})]is a random quantity,
domains of size $\xi$ choose to be in a state equivalent to $\psis$
or $\barpsi$ by the maximization of $\sum_i h_i^x$. As a result a
net magnetic field in the $x$ direction, $\langle {h_i^x} \rangle =
\langle \delta E \rangle/N \propto \xi^{-3/2}$, is added to the
external transverse field.  As the crossover region is approached
$\xi$ is small and the effective transverse field due to the
off-diagonal dipolar interaction is significant. We thus give a
precise physical origin to the conjecture made in
Ref.~\onlinecite{SS05}.

Our analysis above could equally be done by defining $\delta E$ in
Eq.~(\ref{de2f}) as $\sum_k h_k^z \langle S_k^z \rangle$, where $h_k^z
\equiv (2 s \mub H_t/\Omega_o) \sum_i V_{ki}^{zx}$.  Using this
definition one can make the analogy between the current problem to the
Ising SG in random longitudinal field, as an alternative to the direct
calculation of $\delta E$ performed above.

{\it Experimental consequences.---}
%The $\LHx$ system, despite
%having a crystal field Hamiltonian different from the one given in
%Eq.~(\ref{genH}), fulfils the
%
The crystal field Hamiltonian in $\LHx$ is different from the one
given in Eq.~(\ref{genH}). Furthermore, the strong hyperfine
interactions strongly re-normalize the parameters of the TFIM,
invalidating the simple model in the electronic degrees of
freedom\cite{SS05}
%
%invalidate the TFIM in the electronic degrees of freedom, and result
%in an electro-nuclear Ising-like single spin ground state, and a low
%energy effective Hamiltonian given by the TFIM with strongly
%re-normalized parameters~\cite{SS05}.
%
%the combination of the crystal field and the strong hyperfine
%interaction leads to
%
Still, for our purpose here an equivalent physical picture emerges:
the two relevant (electro-nuclear) Ising states of each Ho ion
couple very weakly at small transverse field, and the relevant
excited states are at $\approx 10$K above the ground states
%
%(these two excited states have the same electronic states but differ
%in their nuclear spin~\cite{SS05}).
%
Thus, the requirements for the validity of our theory given before
Eq.~(\ref{genH}) are fulfilled. Our analysis and results [and in
particular Eq.~(\ref{corl})] are therefore directly applicable to
the SG experiments in the $\LHx$ system~\cite{WER+91,WBRA93}, with
$\Omega_o \approx 10$K and suggest that $\LHx$ is not a SG at any
$H_t \neq 0$. Furthermore, the peculiar experimental
result~\cite{WBRA93} where the cusp in the nonlinear susceptibility
is reduced with decreasing $T$ is naturally explained: as $T$ is
reduced the crossover to the PM phase occurs at larger transverse
fields. This results in smaller correlation length $\xi$, and
therefore a diminishing of the cusp in the nonlinear
susceptibility\cite{WBRA93}. In addition, the re-normalization of
the effective spin \cite{SS05} specific to the $\LHx$ compound
further reduces the non-linear susceptibility near the crossover.

From the experimental point of view our analysis changes the status
of the field. The only claim for the observation of the QPT between
the SG and PM phases was made in Ref.\onlinecite{WBRA93}. As it is
clear by our analysis here that it is a crossover and not a phase
transition that was observed at low temperatures in the above
experiment, an experimental observation of this QPT is still
awaiting. Our analysis also points to the direction one should take
in seeking such a QPT: systems in which SG order and quantum
fluctuations compete, and either or both are controlable by a
parameter which does not change the symmetry responsible for the GS
degeneracy of the ordered state. An example would be the change,
with applied pressure, of crystal field terms which induce quantum
fluctuations between the Ising like doublet (such as $(S_+^2 +
S_-^2)$ terms added to the Hamiltonian (\ref{genH}) for integer spin
systems).

Recently there is an increasing
experimental~\cite{MJN+95,JTKI05,SS05b} and
numerical~\cite{TH04,YK04} support for the validity of the droplet
picture in describing short range Ising SG in general, and to its
prediction~\cite{FH86,FH87} of the non-existence of a de
Almeida-Thouless~\cite{AT78} in particular.  For the anisotropic
dipolar systems discussed here the crossover to the PM phase at $H_t
\approx \Omega_o/\mub$ is a result of quantum fluctuations, and
there is no analog to the de Almeida-Thouless line. However, at $H_t
\ll \Omega_o/\mub$ the system is equivalent to a classical Ising SG
in a small random longitudinal field.  Thus, the above numerical and
experimental results \cite{MJN+95,JTKI05,SS05b,TH04,YK04} support
the validity of the droplet picture for dipolar Ising systems in
small transverse field as well. Still, we believe that experiments
that would directly observe whether dipolar Ising glasses in general
and $\LHx$ in particular have a SG phase at a finite transverse
magnetic field are of much interest, both as a verification of our
results, and as an additional support for the droplet picture in
general.

Finally, our analysis is applicable also to any Ising SG where the
dipolar interactions are present, even if the interaction that
governs the ordering is different. In such case the correlation
length will be given by~\cite{SL06} $\xi_J \approx
\left(\frac{\Omega_o J}{\mub H_t V}\right)^\frac{1}{(3/2) -
\theta}$, where $J$ is the strength of the dominant interaction. The
qualitative picture will remain similar, only now the size of the
domains at the quantum crossover to the PM phase would be $\approx
\left(\frac{J}{V}\right)^{1/(3/2 - \theta)}$.

It is a pleasure to thank Ian Affleck, Amnon Aharony, and Philip Stamp
for useful discussion. This work was supported by NSERC of Canada and
PITP. The numerical simulations were carried out on the WestGrid
network, funded in part by the Canada Foundation for Innovation.


\begin{thebibliography}{10}
\bibitem{note2}Computationnaly speaking, the situation here is extreme in the
sense that the spin Hamiltonian (\ref{s1}) on the randomly diluted
$\LHx$ lattice has lost all possible space and internal-spin
symmetries.
\bibitem{note1} Note a finite size effect due to a
  short range interaction which is visible for $N=6$ but disappears at
  larger $N$.
\end{thebibliography}
\end{document}